\documentstyle[11pt,newpasp,twoside,graphicx]{article}
\def\edcomment#1{\iffalse\marginpar{\raggedright\sl#1\/}\else\relax\fi}
\reversemarginpar

\begin{document}
\title{The Extended Methanol Maser Emission in W51}
 \author{Chris Phillips}
\affil{Australia Telescope National Facility, CSIRO, PO Box 76, Epping NSW 1710, 
  Australia}
\author{Huib Jan van Langevelde}
\affil{Joint Institute for VLBI in Europe, 7990 AA Dwingeloo, The Netherlands }

\begin{abstract}
  The European VLBI Network (EVN) has been used to make phase
  referenced, wide-field (several arcminute) spectral line
  observations of the 6.7-GHz methanol maser emission towards W51. In
  the W51main region, the bulk of the methanol is offset from an
  UCH{\sc ii} region. This probably indicates the methanol emission
  arises at the interface of the expanding UCH{\sc ii} region and not
  from an edge-on circumstellar disc, as previously suggested. Near
  the W51 IRS2 region, the methanol emission is associated with a very
  compact, extremely embedded source supporting the hypothesis that
  methanol masers trace the earliest stages of massive star
  formation. As well as these two previously well studied sites of
  star formation, many previously unknown regions star formation are
  detected, demonstrating that methanol masers are powerful means of
  detection young massive stars.

\end{abstract}

\section{Introduction}

Massive stars play a crucial role in the dynamics and evolution of
galaxies but their formation is poorly understood because they are
rare, evolve rapidly and are heavily embedded. There is increasing
evidence that class II methanol masers are associated with with some
of the earliest stages of massive star formation. If this is the case,
methanol masers can be a powerful probe for finding and studying young
massive stars and proto-stars.

W51 is one of the most luminous regions of massive star formation in
our Galaxy, and is at a distance of 7~kpc. The W51a field is known to
show methanol maser emission and two isolated sources, W51~Main and
W51~IRS2, have previously been observed with the EVN at 6.7~GHz
(Phillips, 2000) and at 12.2~GHz with the VLBA (Minier et~al.\
2000). The region was also observed as part of a survey of Northern
methanol maser sources with the ATCA. These observations showed that
the VLBI imaging had missed a large fraction of the maser emission
which was located in 9 separate sites spread over more than
4~arcminutes.

\section{Observations}

The Northern position of W51 resulted in limited {\em uv} coverage and
observations at low elevation, so good astrometry could not be
achieved with the ATCA data. To get high precision absolute and
relative astrometry of the maser emission, we used the EVN in phase
referenced mode to make large field of view observation of the
region. The observations were made on 7 Feb 2002, using 4 EVN antennas
which can observe at 6.7~GHz. The data were correlated with the EVN
MarkIV data processor at JIVE. Because the maser emission is spread
over a region larger than the primary beam of the Effelsberg
telescope, the observations had to be made using two separate
pointings.

\section{Results and Discussion}

\begin{figure}
  \includegraphics[width=0.95\textwidth]{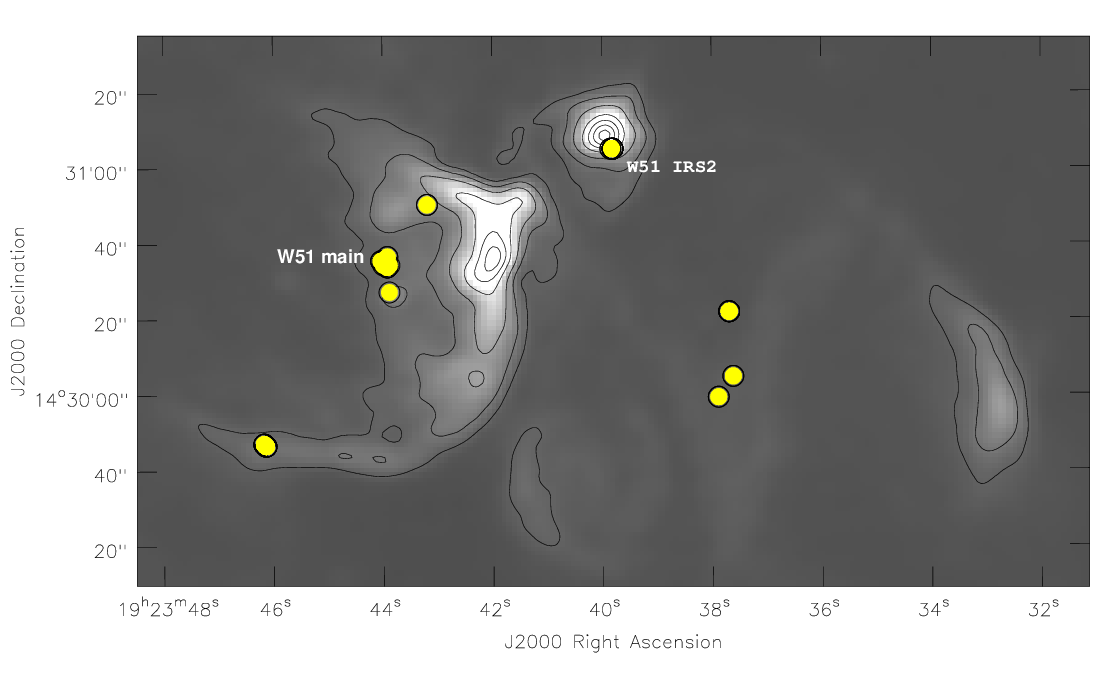}
  \caption{The W51a region observed with the EVN. The white/yellow
    circles with a black outline indicate the position of the 6.7-GHz
    methanol masers, while the greyscale and contours show the 4.8~GHz
    continuum observed with the VLA (Mehringer 1994). }
\end{figure}

Figure~1 shows the measured position of the methanol masers overlayed
on an 4.8-GHz VLA continuum image of the region. Although a few
sources are associated with known centres of activity, the majority of
the masers do not show any obviously compact radio continuum. However,
the presence of the methanol masers means that these are the locations
of some sort of activity, presumably massive star formation.
Interestingly, 800 $\mu$m observations made by Ladd et~al.\ (1993)
using the JCMT, show a compact continuum source at the same position
of the masers in the South-East. The whole region has been surveyed
for main line OH maser emission using the VLA (Argon et~al.\ 2000) and
most of sites of methanol maser emission do not show any OH maser
emission either. VLA observations of H$_2$O masers and thermal NH$_3$
do not detect any emission towards most of the methanol masers, but
given the relatively small primary beam at 22~GHz, this is probably
not surprising without targeted observations.

We suggest that the newly detected methanol masers represent either
very young or very embedded (possibly both) massive stars or
proto-stars. Follow-up observations using instruments such as BIMA and
the VLA at 20 GHz will be needed to understand the origin of these
sources.

Two of the sites of methanol maser emission in this region are well
known sources and have been studies extensively at many wavelength;
W51 Main and W51 IRS2.

\begin{figure}
  \includegraphics[width=0.95\textwidth]{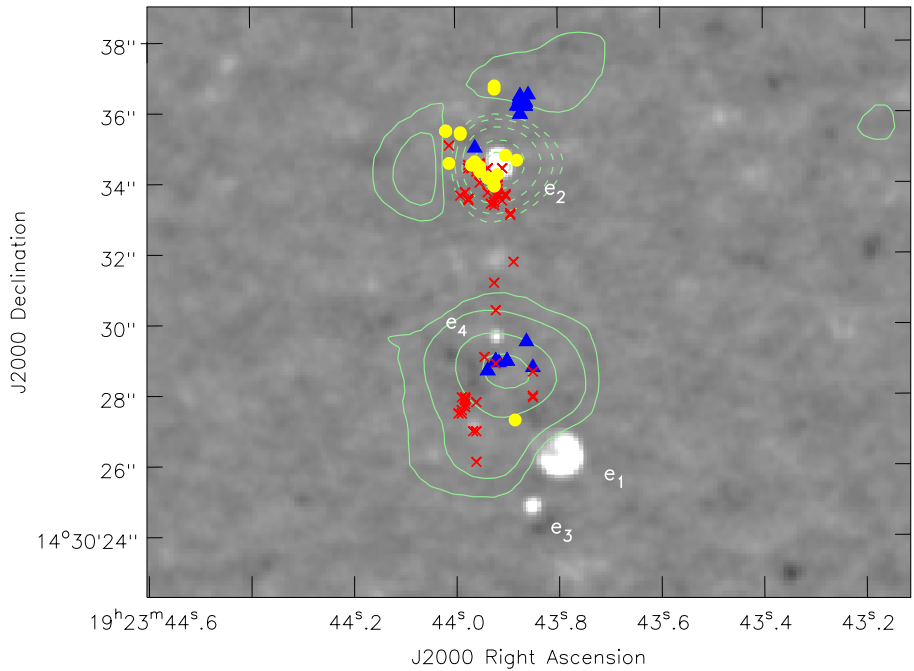}
  \caption{Detail of the W51 Main region. The white/yellow circles
  indicated the position of 6.7-GHz methanol masers, the position of
  OH and water masers are indicated by (red) crosses and (blue)
  triangles (Argon et~al.\ 2000; Imai et~al.\ 2002). The greyscale
  shows the 8.4~GHz continuum emission observed using the VLA A array
  (Gaume et~al.\ 1993) while the (green) contours show the integrated
  NH$_3$ (1,1) emission also observed with the VLA (Zhang \& Ho
  1997).}
\end{figure}

\subsection{W51 Main}

Figure~2 shows a close up of the W51 main region, with the methanol
positions from the current data overlayed on 8.4~GHz continuum
emission as well as OH and water masers and NH$_3$ (1,1) emission. The
astrometry shows that the bulk of the methanol emission is offset to
the edge of a bright, unresolved, UCH{\sc ii} region which shows deep
NH$_3$ absorption, indicating that the object is still embedded in
dense molecular material. Most of the methanol components lie in an
elongated structure which has previously been interpreted as the
masers delineating an edge on disc around a young massive star
(Phillips et~al.\ 2000). Linear structure of the methanol emission is
seen in many other methanol maser sources, and the disc hypothesis is
often used to explain the morphology and velocity gradient along these
sources (Phillips et~al.\ 1998; Norris et~al.\ 1998).  The relative
position of the methanol emission with respects to the UCH{\sc ii}
region found by the current observations, and the fact that the
methanol is highly mixed with the OH emission, suggest that in fact
the methanol emission in this source is not associated with a disc,
but is at the interface of the expanding H{\sc ii} region and the
molecular envelope.

\begin{figure}
  \includegraphics[width=0.95\textwidth]{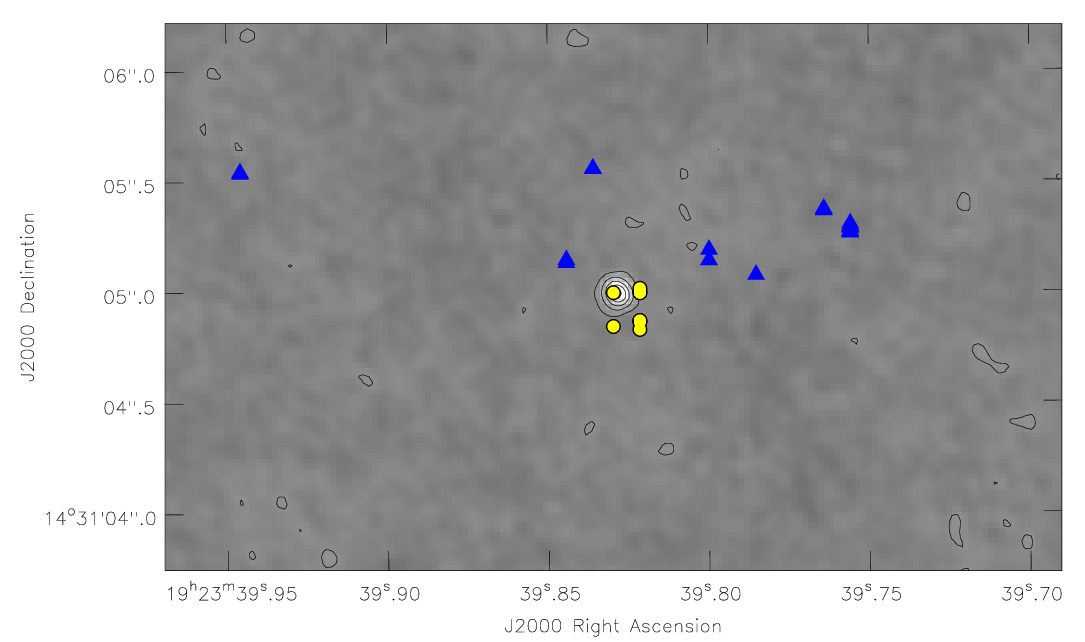}
  \caption{Methanol masers towards the W51 IRS2 region (IRS2 is
slightly to the east of viewable region). Position of methanol masers
are shown by white/yellow circles with a black border.  Black/blue
triangles indicated the position of H$_2$O masers (Eisner et~al.\
2002). The greyscale and contours show the 22 GHz continuum observed
with the VLA-A array (Gaume et~al.\ 1993).}
\end{figure}

\subsection{W51 IRS2}

Figure~3 shows the methanol emission towards W51~IRS2. The methanol
masers are not associated with the infrared source but are offset to
the South-West and are associated with an isolated UCH{\sc ii} region
(Gaume et~al.  1993). The H{\sc ii} region has a spectral index of is
2.1, a physical size $<$300 AU and its emission measure infers a
central star of spectral type of B0.5. This shows that the methanol is
associated with a very young, highly embedded massive
star. Interestingly, the UCH{\sc ii} region is also associated with an
({\em J,~K})~=~(9,~8) NH$_3$ maser. It seems likely that the newly
detected methanol masers are associated with objects similar to this
one with, and either have a lower mass or are younger so no UCH{\sc
ii} region is present or detectable.


\begin{references}

  \reference {Argon}, A., {Reid}, M. \& {Menten}, K. 2000, ApJSS, 129, 159
  \reference {Eisner}, J., {Greenhill}, L., {Herrnstein}, J., {Moran}, J.,
             \& {Menten}, K.  2002, ApJ, 569, 334
  \reference {Gaume}, R., {Johnston}, K. \& {Wilson}, T. 1993, ApJ, 417, 645

  \reference {Imai}, H., {Watanabe}, T., {Omodaka}, T., {Nishio}, M., 
	{Kameya}, O., {Miyaji}, T. \& {Nakajima}, J. 2002, PASJ, 54, 741

  \reference Ladd, E., Deane, J., Goldader, J., Sanders, D., \&
  Wynn-Williams, C. 1993 in AIP Conf. Proc., 278, Back to the Galaxy,
  ed. S. Hold \& F. Verter (New~York, AIP), 246
  \reference Mehringer, D. 1994, ApJSS, 91, 713
  \reference {Minier}, V., {Booth}, R., \& {Conway}, J. 2000, A\&A, 362, 1093
  \reference {Norris}, R., {Byleveld}, S., {Diamond}, P., {Ellingsen}, S., 
	     {Ferris}, R., {Gough}, R.\ G. and 
        {Kesteven}, M.\ J. and {McCulloch}, P.\ M. and {Phillips}, C.\ J. and 
        {Reynolds}, J.\ E. and {Tzioumis}, A.\ K. and {Takahashi}, Y. and 
        {Troup}, E.\ R. \& {Wellington}, K.\ J. 1998, ApJ, 508, 275
  \reference {Phillips}, C., {Norris}, R., {Ellingsen}, S. \& 
      {McCulloch}, P. 1998, MNRAS, 300, 1131
  \reference Phillips 2000, Proceedings of the 5th EVN Symposium
  \reference {Zhang}, Q. \& {Ho}, P. 1997, ApJ, 241
\end{references}
\end{document}